\documentclass[sigconf]{acmart}
\usepackage{amsfonts,amssymb}
\usepackage{amsmath,amssymb}
\newcommand{\ignore}[1]{}
\usepackage{xspace}
\usepackage{bm}

\newcommand{\ie}{\emph{i.e.,}\xspace}

\newcommand{\eg}{\emph{e.g.,}\xspace}

\AtBeginDocument{%
  \providecommand\BibTeX{{%
    \normalfont B\kern-0.5em{\scshape i\kern-0.25em b}\kern-0.8em\TeX}}}

\copyrightyear{2020}
\acmYear{2020}
\setcopyright{acmcopyright}
\acmConference[CIKM '20] {The 29th ACM International Conference on Information and Knowledge Management}{October 19--23, 2020}{Virtual Event, Ireland}
\acmBooktitle{The 29th ACM International Conference on Information and Knowledge Management (CIKM '20), October 19--23, 2020, Virtual Event, Ireland}
\acmPrice{15.00}
\acmDOI{10.1145/3340531.3412098}
\acmISBN{978-1-4503-6859-9/20/10}
\begin{document}
\fancyhead{}

%% Use this when submitting an article to a sponsored event. You'll
%% receive a unique submission ID from the organizers
%% of the event, and this ID should be used as the parameter to this command.
%%\acmSubmissionID{123-A56-BU3}

%%
%% The majority of ACM publications use numbered citations and
%% references.  The command \citestyle{authoryear} switches to the
%% If you are preparing content for an event
%% sponsored by ACM SIGGRAPH, you must use the "author year" style of
%% the next command will enable that style.

%%
%% end of the preamble, start of the body of the document source.

\title{Leveraging Historical Interaction Data for Improving Conversational Recommender System}

\author{Kun Zhou$^{1}$, Wayne Xin Zhao$^{2,3*}$, Hui Wang$^{1}$, Sirui Wang$^{4}$, Fuzheng Zhang$^{4}$, \and Zhongyuan Wang$^{4}$ and Ji-Rong Wen$^{2,3}$}\thanks{$^*$Corresponding author.}
\affiliation{%
 \institution{$^1$School of Information, Renmin University of China}
 \institution{$^2$Gaoling School of Artificial Intelligence, Renmin University of China}
 \institution{$^3$Beijing Key Laboratory of Big Data Management and Analysis Methods}
 \institution{$^4$Meituan-Dianping Group}
}
\affiliation{%
  \institution{francis\_kun\_zhou@163.com, \{batmanfly, hui.wang, jrwen\}@ruc.edu.cn, \and wangsirui@meituan.com, zhfzhkris@outlook.com}
}

\begin{abstract}
Recently,
conversational recommender system (CRS) has become an emerging and practical research topic. Most of the existing CRS methods focus on learning effective preference representations for users from conversation data alone. While, we take a new perspective to leverage historical interaction data for improving CRS. 
For this purpose, we propose a novel pre-training approach to integrating both item-based preference sequence (from historical interaction data) and attribute-based preference sequence (from conversation data) via pre-training methods.
We carefully design two pre-training tasks to enhance information fusion between item- and  attribute-based preference. To improve the learning performance, we further develop an effective negative sample generator which can produce high-quality negative samples. Experiment results on two real-world datasets have demonstrated the effectiveness of our approach for improving CRS. 
\end{abstract}

%\ccsdesc[300]{Information systems~Personalization}


\maketitle
\section{Introduction}
With the rapid development of intelligent agents in e-commerce platforms, 
conversational recommender system (CRS)~\cite{DBLP:conf/sigir/SunZ18,Chen2019TowardsKR,DBLP:conf/wsdm/Lei0MWHKC20}
has become an emerging research topic in seeking to provide high-quality recommendations to users through conversations.
Generally, a CRS consists of a conversation module and a recommendation module.
The conversation module focuses on acquiring users' preference via multi-turn interaction, and the recommendation module focuses on how to utilize the inferred preference information to recommend suitable items for users.

Most of the existing CRSs are designed in a ``\emph{system asks-user responds}'' mode~\cite{DBLP:conf/wsdm/Lei0MWHKC20,DBLP:conf/sigir/SunZ18}. At each round, CRS issues a query about user preference and the user replies to the system with personalized feedback.
Typically, a system query is generated according to some attributes of items (\eg \emph{what is your favorite movie genre}), and the user feedback
reflects the specific preference of a user on that attribute (\eg \emph{I like Action movies}). A mainstream approach~\cite{DBLP:conf/sigir/SunZ18,DBLP:conf/wsdm/Lei0MWHKC20} is to construct a belief tracker module that infers the attribute-based preference of a user from such a multi-turn conversation. 
In this way, the inferred preference can be presented as a sequence of inferred attributes (\eg ``\textsc{Genre}=\emph{Action}  $\rightarrow$ \textsc{Director}=\emph{James Cameron}'' in a movie CRS). Furthermore, Factorization Machine~\cite{DBLP:conf/sigir/SunZ18,DBLP:conf/wsdm/Lei0MWHKC20} or KG-based models~\cite{Chen2019TowardsKR} can be applied to construct the user preference representation and make the final recommendation. 

However, these existing studies for CRS suffer from two major issues. First, the information of a conversation itself is quite limited. Many CRSs are further optimized to reduce the number of rounds that the system interacts with users~\cite{DBLP:conf/wsdm/Lei0MWHKC20,DBLP:conf/sigir/SunZ18}. Thus, some useful attributes are likely to be missed in the inferred attribute-based preference.
Second, it may be not sufficient to utilize attribute-based preference alone for making the recommendation. For example, the candidate item set can be still large even after the filter of several attributes. 

To address the two issues, 
%we observe an important phenomenon: a CRS is usually deployed within an application platform. 
we observe that a CRS is usually deployed within an application platform. 
When an existing user from the platform enters into its deployed CRS, we can obtain his/her historical interaction data, \ie a chronologically-ordered item sequence of the user. Intuitively, historical interaction data provides another kind of useful data signal to infer user preference. However, it is not easy to integrate the two kinds of information for CRS, which are different in essence.  
Indeed, historical interaction reflects item-level user preference in long run, while conversation data reflects attribute-level user preference at present. Therefore, it is important to develop an effective model that can fuse item-level and attribute-level user preference for CRS.

Inspired by the success of pre-training methods like BERT~\cite{DBLP:conf/naacl/DevlinCLT19}, we propose a novel pre-training approach to leveraging historical interaction data of users for improving conversational recommendation. 
Our main idea is to integrate both item-based preference sequence (from historical interaction data) and attribute-based preference sequence (from conversation data) via pre-training methods. 
To model the two kinds of preference sequences, we develop an approach based on a self-attentive architecture ~\cite{DBLP:conf/icdm/KangM18}, containing an item-based Transformer and an attribute-based Transformer.
Specially, we design two auxiliary tasks for enhancing the data fusion, namely 
\emph{Masked Item Prediction~(MIP)} and \emph{Substituted Attribute Discrimination~(SAD)}. The MIP strategy adapts the idea of the Masked Language Model (MLM) in BERT~\cite{DBLP:conf/naacl/DevlinCLT19} to conversational recommendation; the SAD strategy emphasizes the capacity of discriminating between actual or substituted attributes based on contextualized item representations. To improve the pre-training performance, we further propose to generate high-quality negative samples with a pre-training model by following IR-GAN~\cite{DBLP:conf/sigir/WangYZGXWZZ17} and ELECTRA~\cite{DBLP:journals/corr/abs-2003-10555}.

%\ignore{Our contributions are summarized as follows: 
To our knowledge, it is the first time 
that historical interaction data has been integrated and utilized in a setting of CRS by a pre-training approach. 
We believe such an idea is promising to improve existing CRS methods. 
To demonstrate the effectiveness of our approach, we construct experiments on real-world datasets that are tailored to the CRS task. 
Experiments show that our approach is more effective than a number of competitive CRS methods. 

\section{Problem Statement}
\begin{figure}
\includegraphics[width=0.45\textwidth]{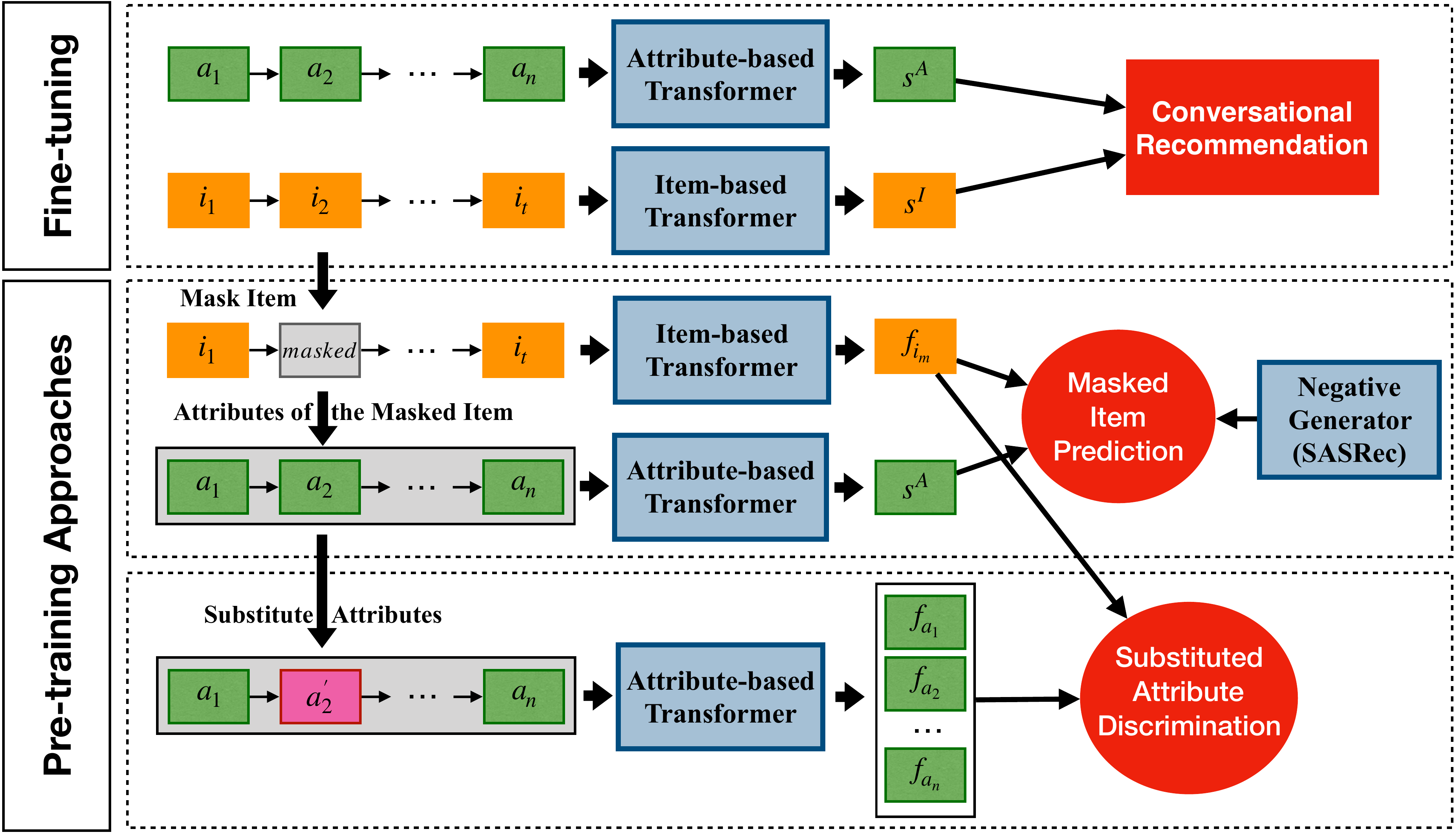}
\caption{The overview of our approach. During pre-training, we incorporate MIP and SAD tasks to optimize the parameters. 
During fine-tuning, we optimize our pre-trained model by conversational recommendation task.}
\label{fig-approach}
\end{figure}

When a user $u$ enters into a CRS, a multi-round conversation will be initiated for making an accurate recommendation in a ``\emph{system asks-user responds}'' mode.  
As discussed before, most of existing CRSs~\cite{DBLP:conf/wsdm/Lei0MWHKC20,DBLP:conf/sigir/SunZ18} focus on learning attribute-level user preference for user preference. 
Typically, the queries from CRS are about the user preference over the possible item attributes, which is denoted by an attribute set $\mathcal{A}$. 

As the conversation goes on,  the system becomes increasingly clear about user preference, since it has acquired an \emph{attribute-based preference sequence} from the target user, denoted by $\mathcal{P}^{(A)}=a_1\rightarrow \cdots  a_j \cdots \rightarrow a_n$, where $n$ is the sequence length and $a_j$ belongs to the attribute set $\mathcal{A}$.
For a movie CRS, an example for the obtained $\mathcal{P}^{(A)}$ can be given as: \textsc{Genre}=\emph{Action}  $\rightarrow$ \textsc{Director}=\emph{James Cameron}.

Besides, we assume that the interaction history of user $u$ is also available in our setting.  It is given as an \emph{item-based preference sequence}, denoted by $\mathcal{P}^{(I)}=i_1\rightarrow \cdots  i_k  \cdots \rightarrow i_m$, where each $i_k \in \mathcal{I}$ is an item that user $u$ has interacted with at the $k$th step before the conversation, and $\mathcal{I}$ is the item set. We further assume that each item $i_k$ is also associated with a set of attribute values, denoted by $\mathcal{A}_{i_k}$, which is a subset of the entire attribute set $\mathcal{A}$.

Based on these notations, the task in our paper is defined as: given both attribute- and item-based preference sequences $\mathcal{P}^{(A)}$ and  $\mathcal{P}^{(I)}$, we aim to accurately predict the item $i^{*}$ that well matches the needs of user $u$ in a conversation. 

Note that our task setting is slightly different from existing CRS studies~\cite{DBLP:conf/sigir/SunZ18,DBLP:conf/wsdm/Lei0MWHKC20}. Here, our focus is not to ask good questions or form suitable text-based responses. 
We aim to improve the recommendation task using historical interaction data after already acquiring some attribute preference information from the user.
Therefore, we assume that the attribute-based preference $\mathcal{P}^{(A)}$ can be obtained through existing methods (\ie belief tracker~\cite{DBLP:conf/sigir/SunZ18,DBLP:conf/wsdm/Lei0MWHKC20}). 
Indeed, such a task setting can be considered as a sub-task or sub-module for existing CRSs, \ie how to effectively generate the item recommendations based on \textit{interaction history} and \textit{conversation data}. In practice, CRS can run such a module for our task multiple times once the acquired attribute preference is updated. 
%It is noted that our task formulation is also related to faceted search~\cite{DBLP:conf/icde/VandicAFK18} and context-aware sequential recommendation~\cite{DBLP:conf/icdm/LiuWWLW16}. 
%Currently, we frame the task in the setting of CRS, and we will consider investigating how the similar ideas apply to the above related tasks in future works. 
\section{Approach}
In this section, we present the proposed approach to the item recommendation task in CRS, which is inspired by the recently proposed pre-trained language models~\cite{DBLP:conf/naacl/DevlinCLT19}.
Our approach contains two major stages, namely the pre-training and fine-tuning stages. We aim to learn effective representations through information fusion in the pre-training stage. Then we fine-tune the model according to the CRS task. 
%We adopt the Transformer architecture~\cite{vaswani2017attention} in both stages, which has been shown effective in various applications.
The major contribution of our work lies in the pre-training stage, where we carefully design several auxiliary pre-training tasks to fuse item- and attribute-based preference. We present an overview of our approach in Fig.~\ref{fig-approach}.

\subsection{Base Model}
\label{base}
We adopt Transformer~\cite{vaswani2017attention} as our base model, which consists of embedding layer, self-attention layer, and prediction layer.

In the embedding layer, we maintain an item embedding matrix $\mathbf{M}_{I}\in \mathbb{R}^{|\mathcal{I}|\times d}$ and an attribute embedding matrix $\mathbf{M}_{A}\in \mathbb{R}^{|\mathcal{A}|\times d}$.
The two matrices project the high-dimensional one-hot representation of an item or attribute to low-dimensional dense representation. 
Furthermore, we incorporate a learnable position encoding matrix $\mathbf{P}\in \mathbb{R}^{n\times d}$ to enhance the input representations of item sequence.

Based on the embedding layer, we build item and attribute encoders by stacking multiple self-attention blocks, and the two encoders are implemented with the same architecture but with different parameters. A self-attention block generally consists of two sub-layers: a multi-head self-attention layer and a point-wise feed-forward network. More details can be found in~\cite{vaswani2017attention}.

In the prediction layer of our model, we compute the preference score of user $u$ for each candidate item $i$ based on the interaction history $\mathcal{P}^{(I)}$ and conversation data $\mathcal{P}^{(A)}$ as:
\begin{eqnarray}
\label{eq-ft}
    P(i | u, \mathcal{P}^{(A)}, \mathcal{P}^{(I)} ) \propto [\bm{s^{I} ;s^{A}} ]^{\top} \mathbf{W}_{M} \bm{e}_i, 
\end{eqnarray}
where $\bm{e}_i$ is the representation of item $i$ from item embedding matrix $\mathbf{M}_{I}$,
\bm{$s^{I}$} and \bm{$s^{A}$} are the learned state representations of the last position from the item and attribute encoders respectively, and $\mathbf{W}_{M}\in \mathbb{R}^{2d\times d}$ is the trainable parameter matrix. 

\subsection{Improvement with Pre-training Strategies}
\label{pre-train}
Based on the above model architecture, 
we propose two pre-training tasks, the \emph{Masked Item Prediction} task~(MIP) and the \emph{Substituted Attribute Discrimination} task~(SAD), to enhance the data representations via effective fusion of item- and attribute-based preference. 

\subsubsection{Masked Item Prediction}
Inspired by BERT~\cite{DBLP:conf/naacl/DevlinCLT19}, we construct a \emph{Cloze} task based on the item sequence.
Given an item sequence $\mathcal{C}=i_1\rightarrow \cdots  i_k  \cdots \rightarrow i_m$, we randomly mask a proportion of items in the sequence, \eg replacing with special token ``[\emph{mask}]''.
For each masked item, we predict its original ID based on its contextual information, consisting of contextual items $\mathcal{C}_{\neg i_k}$ and attributes $ \mathcal{A}_{i_k} $. 
Following BERT~\cite{DBLP:conf/naacl/DevlinCLT19}, we leverage the bidirectional contextual information in item sequence for predicting the masked item as: 
\begin{eqnarray}
    P(i_k | \mathcal{C}_{\neg i_k}, \mathcal{A}_{i_k} )=\sigma \big([\bm{f_{i_k};s^{A}} ]^{\top} \mathbf{W}_{M} \bm{e}_{i_k}\big)
\end{eqnarray}
where $\bm{f}_{i_k}$ is the representations for $\mathcal{C}_{\neg i_k}$, which is the representation for the $k$-th position using the bidirectional Transformer.
$\bm{s^{A}}$ is the representation of $\mathcal{A}_{i_k}$ as in Eq.~\ref{eq-ft}, and $\sigma (.)$ is the sigmoid function to obtain the probability.
We adopt the pairwise ranking loss with negative sampling to learn this pre-training task.

\subsubsection{Substituted Attribute Discrimination}
Furthermore, we design another task that enhances the fusion between item- and attribute-level information.
In an item sequence, let $\mathcal{A}_{i_k}$ denote the associated attributes for the item $i_k$.
Following ELECTRA~\cite{DBLP:journals/corr/abs-2003-10555} which replaces word in sentence by another irrelevant word,
we randomly substitute some of its attributes in $\mathcal{A}_{i_k}$ with negative attributes that are randomly sampled, and obtain a corrupted $\tilde{\mathcal{A}}_{i_k}$. 
In this way, our second \emph{Cloze} task to predict if the attribute $a_j$ in $\tilde{\mathcal{A}}_{i_k}$ has been replaced or real:
\begin{eqnarray}
    P(y_{a_j}=1 | \mathcal{C}, \tilde{\mathcal{A}}_{i_k} )=\sigma\big(\bm{f}_{i_k} ^{\top} \mathbf{W}_{P} \bm{f}_{a_j}\big),
\end{eqnarray}
where $y_{a_j}$ is a discrimination label for $a_{j}$, $\bm{f}_{i_k}$ and $\bm{f}_{a_j}$ are the representations of item $i_k$ and attribute $a_j$ through our architecture, respectively, and $W_{P}\in \mathcal{R}^{d\times d}$ is the trainable parameter matrix.
We adopt the cross-entropy loss to learn this pre-training task.

\subsection{Learning with Enhanced Negative Samples}
\label{negative}
A key procedure to learn with the MIP task is to sample irrelevant items as \emph{negative samples}. 
In recommender systems~\cite{DBLP:conf/icdm/KangM18,DBLP:conf/sigir/SunZ18,DBLP:conf/icdm/LiuWWLW16}, the most commonly adopted way is to randomly sample negative items.
However, it has been widely recognized that the quality of negative samples directly affects the models~\cite{DBLP:journals/corr/abs-2003-10555,DBLP:conf/sigir/WangYZGXWZZ17,DBLP:conf/wsdm/Lei0MWHKC20}: too easy or too difficult negative samples are likely to lead to worse performance. 

Inspired by IR-GAN~\cite{DBLP:conf/sigir/WangYZGXWZZ17} and ELECTRA~\cite{DBLP:journals/corr/abs-2003-10555}, we propose a new strategy to derive negative samples for the masked item prediction task,
%Similar to the idea of Generative Adversarial Network, 
in which we set up a special module for generating high-quality negative samples. 
Since our setting is in a sequential manner, we utilize the state-of-the-art sequential recommendation model (\ie SASRec~\cite{DBLP:conf/icdm/KangM18}) as a negative sample generator.  
We first pre-train the generator with traditional pairwise ranking loss using random sampling. Then, we sample negative samples according to the probability distribution that it assigns to each candidate item. 
Since SASRec has achieved very good performance in sequential recommendation, the top-ranked items with high probabilities are likely to be ``\emph{close-to-real}'' ones, which is helpful to improve the learning of our approach. Note that although we can update the generator as that in standard GAN, we train its parameters only once. We have empirically found that the improvement with iterative updating is limited in our task.

Finally, the entire training procedure of our approach consists of the pre-training and fine-tuning stages. In the pre-training stage, we apply the two proposed pre-training strategies (with the improved negative sampling method) to enhance the learning of the data representations. In the fine-tuning stage, 
we adopt the pairwise rank loss to re-optimize the parameters according to our task:
\begin{eqnarray}
\label{ft}
  L=  \sum_{c \in \mathcal{D}} \log \sigma \big(P(i^{+}| u, c)-P(i^{-}| u, c) \big),
\end{eqnarray}
where $c$ is a conversation involving user $u$ from training data $\mathcal{D}$, and $i^{+}$ and $i^{-}$ are the actual or negative items in this conversation. 
\section{Experiment}
\subsection{Experimental Setup}
\subsubsection{Datasets and Setup}
\begin{table}
\caption{Statistics of the datasets after preprocessing.}\label{dataset}
%\small
\centering
\begin{tabular}{ccccc}
\hline
    \textbf{Datasets} &\textbf{\#Users} &\textbf{\#Items} &\textbf{\#Interactions} &\textbf{\#Attributes} \\
    \hline
     Meituan &14,290 &30,839 &727,954 &331 \\
    \hline
    LastFM &2,100 &1,921 &39,828 &71 \\
    \hline
  \end{tabular}
\end{table}
We conduct experiments on two datasets: Meituan and LastFM. Meituan dataset contains 6-year (2014-2020) shopping transactions in Beijing on the food subcategory of the Meituan platform\footnote{\url{https://www.meituan.com}}. LastFM dataset is adopted by EAR~\cite{DBLP:conf/wsdm/Lei0MWHKC20}, which is a music artist recommendation dataset and shared in HetRec 2011 tracks\footnote{\url{https://grouplens.org/datasets/hetrec-2011/}}. 
We rebuild these datasets for our task following~\cite{DBLP:conf/wsdm/Lei0MWHKC20}.
The statistics of the two datasets are summarized in Table~\ref{dataset}. 

\subsubsection{Evaluation metrics} 
To evaluate the CRS models, we adopted the \textit{leave-one-out} evaluation~\cite{DBLP:conf/icdm/KangM18,DBLP:conf/icdm/LiuWWLW16}. 
Each user has a sequence of conversational recommendation records, we hold out the last record in user history as the test data, treat the record just before the last as the validation set, and utilize the remaining data for training. 
Besides, each conversational recommendation record contains a targeted item.
We pair the targeted item in the test set with 100 randomly sampled negative items. Hence, the task becomes to rank these items for each user.
We adopt NDCG@10 and MRR to evaluate the performance of the ranking list for all the models.

\subsubsection{Models}
We consider the following baselines for comparisons:
(1) CRM~\cite{DBLP:conf/sigir/SunZ18} utilizes factorization machines to learn the interaction between user, item, and attribute;
(2) EAR~\cite{DBLP:conf/wsdm/Lei0MWHKC20} utilizes linear interactions among user, item, and attribute;
(3) GRU$_{I}$~\cite{DBLP:conf/icdm/LiuWWLW16} adopts Gated Recurrent Units to model user history;
(4) SASRec$_{I}$~\cite{DBLP:conf/icdm/KangM18} adopts self-attentive model to leverage user history;
(5) GRU$_{I+A}$ uses a GRU module to model user history, and a self-attention module to model the conversation data, in which the two types of information are fused by a linear layer;
(6) SASRec$_{I+A}$ adopts two self-attentive modules to model user history and conversation data, respectively, in which the two types of information are fused by a linear layer.

Among all the above methods, CRM and EAR are CRS models, which do not model the sequential pattern in user history. Then, GRU$_{I}$ and SASRec$_{I}$ only utilize the sequence information in user history, while GRU$_{I+A}$ and SASRec$_{I+A}$ utilize both user history and conversation data. Our approach adopts MIP and SAD tasks to pre-train the parameters in our model, in which we further improve MIP task by a negative sample generator (NG). 
The details of our model and dataset are available at this link:
\textcolor{blue}{\url{https://github.com/RUCAIBox/Pre-CRS}}.
%which can fuse the information from user history and conversation data effectively. 
%We further improve our model by a negative generator (NG).
To evaluate the effectiveness of the two pre-training tasks and NG, we conduct an ablation study by removing one component from our approach at each time. 

%\subsubsection{Training Details}
%In our experiments, we implement our model by PyTorch\footnote{https://pytorch.org/}. 
%The batch size is set to 256 and the embedding dimension is set to 64. The mask proportion of items is set to 20\% while the substitution proportion of attributes is set to 50\%.
%The learning rate is set as 0.001 and 0.0001 in the pre-training and fine-tuning stage, respectively. We set the layer number of Transformer to 2 for Meituan dataset, while 4 for LastFM dataset. We use Adam optimizer with its default parameter setting and use gradient clipping to restrict the norm of gradients within [0,0.1]. Our code and data are available at
%\textcolor{blue}{\url{https://github.com/RUCAIBox/Pre-CRS}}

\subsection{Results and Analysis}
\begin{table}
\caption{Performance comparisons of different methods on this task. The number marked with ``*'' indicates the improvement is statistically significant compared with the best baseline (t-test with p-value $< 0.05$). }\label{result}
%\small
\centering
\begin{tabular}{lcccc}
\hline
    Datasets& \multicolumn{2}{c}{Meituan} & \multicolumn{2}{c}{LastFM} \\
    \hline
     Models &MRR &NDCG@10 &MRR &NDCG@10 \\
    \hline
    \texttt{CRM} &0.0942 &0.1009 &0.0567 &0.0547 \\
    \texttt{EAR} &0.0838 &0.0869 &0.0483 &0.0512 \\
    \hline
    \texttt{GRU$_{I}$} &0.0964 &0.1036 &0.0692 &0.0734 \\
    \texttt{SASRec$_{I}$} &0.1001 &0.1083 &0.0783 &0.0778 \\
    \hline
    \texttt{GRU$_{I+A}$} & 0.0825& 0.0847& 0.1034 & 0.1170 \\
    \texttt{SASRec$_{I+A}$} & 0.1327& 0.1496& 0.1231 &0.1295 \\
    \hline
    \textbf{\texttt{Ours}} &\textbf{0.2026}* &\textbf{0.2354}* &\textbf{0.1800}* &\textbf{0.2032}*\\
    \hline
    \texttt{-w/o MIP} &0.1495 &0.1722 &0.0828 &0.0841 \\
    \texttt{-w/o SAD} &0.0627 &0.0604 &0.1091 &0.1216 \\
    \texttt{-w/o NG} &0.1760 &0.2022 &0.1708& 0.1926\\
    \hline
  \end{tabular}
\end{table}
Table~\ref{result} presents the performance of different methods on the conversational recommendation task.
First, we can observe that CRM and EAR do not perform very well in our task setting, since they do not utilize the user history effectively. 
Second, GRU$_{I+A}$ and SASRec$_{I+A}$ perform better than GRU$_{I}$ and SASRec$_{I}$, which indicates that it is useful to leverage the user history and conversation data jointly.
However, the result of GRU$_{I+A}$ is worse than GRU$_{I}$ in Meituan dataset. One possible reason is that the RNN architecture limits the information fusion.
Third, the self-attentive models SASRec$_{I}$ and SASRec$_{I+A}$ perform better than GRU-based models, it indicates that the self-attentive architecture is particularly suitable for the sequential data in this task.
Furthermore, our approach consistently outperforms all the baselines, which indicates the effectiveness of our pre-training method and negative sample generator. Our approach fuses the item- and attribute-level user preference, so that it can improve the performance on conversational recommendation task. Compared with SASRec$_{I+A}$ which has the same architecture as ours, the pre-training method brings large improvement on both datasets. It indicates the effectiveness of our pre-training approach.

Finally, comparing our approach with its ablation variants, we can see that the three components all contribute to the final performance. 
%After removing each of the components, the performance significantly drops.
After removing the Substituted Attributes Discrimination or Masked Item Prediction task, the performance significantly drops. It indicates the importance of the two components.
\section{Conclusion}
This paper presented a novel pre-training approach for conversational recommendation task, which focused on leveraging the item sequence from user history and attribute sequence from conversation data effectively.
Based on a self-attentive architecture, our approach designed two pre-training tasks, namely Masked Item Prediction (MIP) and the Substituted Attributes Discrimination (SAD). We further improved our pre-training method by introducing a negative generator to produce high-quality negative samples.
Experimental results on two datasets demonstrated the effectiveness of our approach for conversational recommendation task. 
As future work, we will investigate how to apply our approch to other related  recommendation tasks, especially context-aware sequential recommendation and faceted search tasks.

\section*{Acknowledgement}
This work was partially supported by the National Natural Science Foundation of China under Grant No. 61872369 and 61832017,  Beijing Academy of Artificial Intelligence (BAAI) under Grant No. BAAI2020ZJ0301, and Beijing Outstanding Young Scientist Program under Grant No. BJJWZYJH012019100020098, the Fundamental Research Funds for the Central Universities, the Research Funds of Renmin University of China under Grant No.18XNLG22 and 19XNQ047. Xin Zhao is the corresponding author.

\bibliographystyle{ACM-Reference-Format}
\bibliography{sample-base}

\appendix

\end{document}